# Video-based analysis of the transition from slipping to rolling

**Álvaro Suárez,** CFE, Uruguay; alsua@outlook.com
**Daniel Baccino,** CFE, Uruguay; dbaccisi@gmail.com
**Arturo C. Martí,** Universidad de la República, Uruguay, marti@fisica.edu.uy

The problem of a disc or cylinder initially rolling with slipping on a surface and subsequently transitioning to rolling without slipping is often cited in textbooks [1-2]. Students struggle to qualitatively understand the difference between kinetic and static frictional forces—*i.e.*, whereas the module of the former is known, that of the latter can only be described in terms of an inequality while the relative velocity at the point(s) of contact is equal to zero. In addition, students have difficulty understanding that frictional forces can act in the direction of motion—*i.e.*, they can accelerate objects [3-6].

Because the time evolution of the linear and angular velocities of a rigid body cannot readily be determined experimentally, the problem is usually addressed from a purely theoretical perspective. One remarkable exception is Ref. [7] where, in a different approach, the authors consider only the evolution of the centre-of-mass velocity of a initially rolling cylinder. Clearly, the lack of experimentation may hamper the learning process, as it does not allow students to visualize and internalize facts that challenge pre-existing misconceptions formed in earlier stages of learning.

The following experiment serves to clearly demonstrate the transition from rolling with slipping to rolling without slipping. In the experiment, a rotating bicycle wheel was placed in contact with a horizontal surface and the wheel in motion was tracked using *Tracker* video analysis software [8]. The software created linear velocity plots for the centre of mass and a point on the circumference as well as a plot of the angular velocity of the rotating wheel. The time evolution plots created by *Tracker* clearly illustrate the transition between the two types of motion.

### Experimental

The experimental set-up consisted of a bicycle wheel of radius R = 32.0 cm initially rotating on its axis at a certain distance above the floor. The wheel was then placed in contact with the floor and freed, as shown in Figure 1. From the moment it came in contact with the floor to the moment it started to rotate without slipping, the wheel in motion was recorded with a Kodak PlaySport video camera mounted on a tripod, with its optical axis perpendicular to the plane of rotation of the wheel.

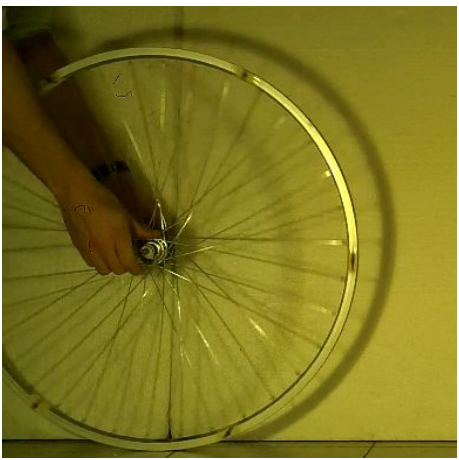

Figure 1. Experimental set-up: A rotating wheel was placed in contact with the floor. The centre of mass and a point located on the wheel rim were tracked and the displacement versus time data were analysed using

*Tracker*.

The results of the analysis are shown in Figures 2 and 3. The plot of the velocity of the centre of mass, $v$, versus time, $t$, shows that, after the wheel was placed in contact with the floor, its centre of mass displayed different types of motion before and after a point in time denoted by $t_1$, when it transitioned between the two types (Figure 2). During the first time interval, the centre of mass of the wheel travelled in uniformly accelerated rectilinear motion, whereas, during the second time interval, starting at approximately $t_1 \sim 0.9$ s, it travelled in uniform rectilinear motion. Results of linear fitting showed that the acceleration of the wheel's centre of mass during the first time interval was $a = (1.31 \pm 0.035)$ m/s² and that its initial velocity was $v_0 = (0.179 \pm 0.019)$ m/s .

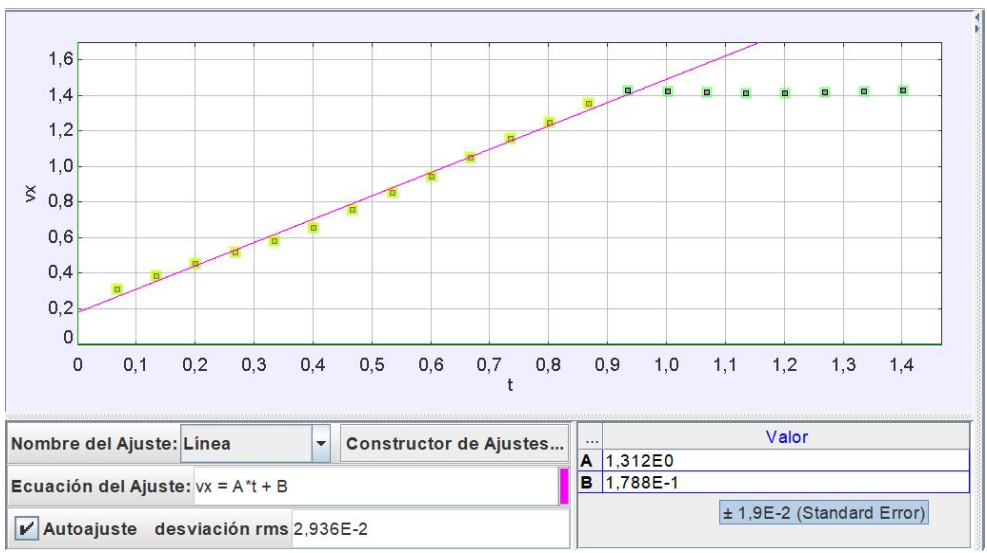

Figure 2. Screenshot of *Tracker* interface showing the time evolution of the velocity of the centre of mass (velocity in m/s and time in s) and the results of linear fitting over the interval in which the wheel moved with constant acceleration. On moving the mouse over the values obtained by linear fitting, a pop-up window showing the corresponding standard error is displayed.

The position of a point on the wheel's circumference—keeping the frame of reference fixed to its centre of mass—was tracked automatically and used to calculate the wheel's angular velocity. Figure 3 shows a plot of the angular velocity, $\omega$, versus time. As in Figure 2, the point displayed two different types of motion over two time intervals. During the first time interval (up to $t_1 \sim 0.9$ s), the module of the angular velocity decreased with time and then the wheel transitioned to uniform circular motion, with its centre of mass as rotation axis. Linear fitting over the first time interval led to an angular acceleration $\alpha = (7.28 \pm 0.24)$ rad/s² and an initial angular velocity $\omega_0 = (-11.4 \pm 0.14)$ rad/s.

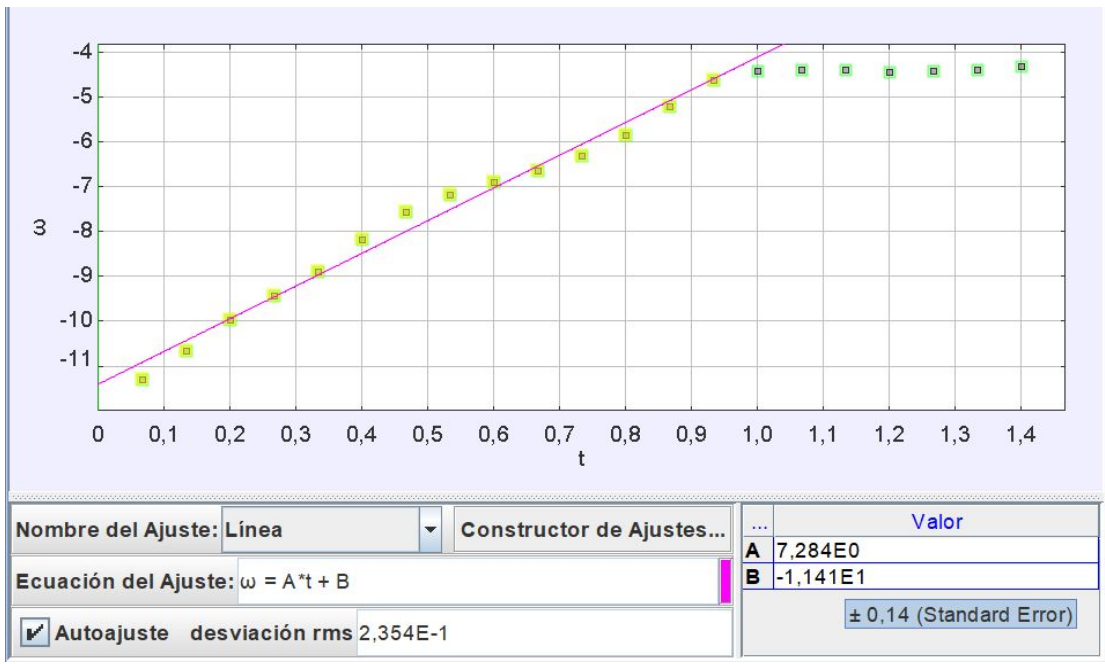

Figure 3. Screenshot of *Tracker* interface showing the time evolution of the angular velocity (in rad/s) and the parameters obtained by linear fitting over the first time interval.

### Discussion and theoretical analysis

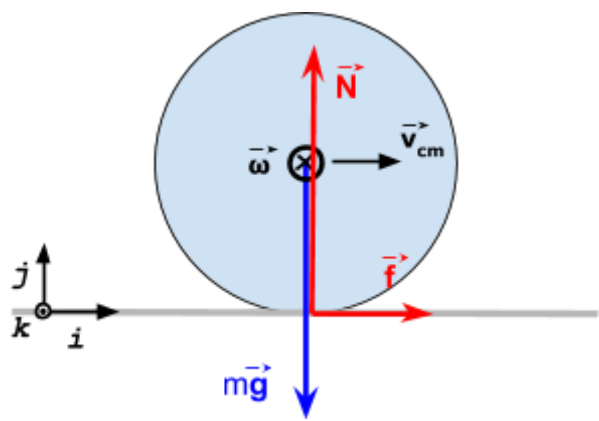

Figure 4. Free body diagram and definition of the coordinate system employed. When the wheel was placed in contact with the floor its angular velocity was $\omega_o$; at this point in time, the centre of mass of the wheel started to move along a straight line.

The forces acting on the wheel from the moment of contact and through the interval where slipping occurs (the first time interval) are shown in Figure 4. In addition to the force of gravity, the wheel is subject to a contact force—i.e., appearing as a result of the interaction between the wheel and the ground—equal to the sum of the normal force $N$ and the kinetic frictional force $f$. Since the vertical acceleration is equal to zero, it follows that

$$N = m.g \qquad (1)$$

$$ma = f \qquad (2)$$

where $a$ is the horizontal acceleration along the x-axis. Meanwhile, the kinetic frictional force is such that

$$f = \mu . N \qquad (3)$$

μ being the friction coefficient between the wheel and the ground. Combining equations (1), (2) and (3) gives the module of the frictional force and the linear acceleration of the centre of mass of the wheel on the x-axis, according to

$$f = \mu . m . g \quad (4)$$
$$a = \mu . g \quad (5)$$

In the experiment, the results of linear fitting, taking $g = 9.8$ m/s², led to

$$\mu = 0.134 \pm 0.004$$

In the interval where slipping occurs, the angular velocity of the wheel cannot be related to the velocity of the centre of mass. Applying Newton's Second Law to the rotation of the wheel around its centre of mass, the only acting torque is generated by the force of friction, according to

$$f . R = I . \alpha \quad (6)$$

where $I$ is the moment of inertia of the wheel about an axis passing by the center of mass, such that

$$I = K m R^2 \quad (7)$$

where $K$ is a dimensionless constant dependent on the distribution of mass, and $\alpha$ the angular acceleration of the wheel. Combining equations (4), (6) and (7) gives

$$\alpha = \frac{\mu g}{K R} \quad (8)$$

In the experiment, given the kinetic friction coefficient, the radius of the wheel and the angular velocity, the distribution of mass of the wheel with respect to the calculated centre of mass can be characterized in terms of $K = 0.56 \pm 0.03$. Slightly above ½, this value indicates that the wheel's mass is largely concentrated on its rim.

After time $t_1$, the wheel displays a pure rolling motion, where no slipping takes place. During this interval, the modules of linear and angular velocities can be related to each other by means of the wheel's radius, as follows

$$v = R . \omega \quad (9)$$

In the experiment, the radius—calculated using the linear and angular velocity data for the second time interval—was $R \sim 0.33$ m, consistent with the measured radius.

It is worth noting that the shapes of the linear acceleration and angular velocity plots showed that the transition from rolling with slipping to rolling without slipping occurred over a short time lapse. Assuming a sharp transition, equation (9) can be combined with equations (10) and (11), below, in order to estimate $t_1$, according to

$$v_1 = v_0 + a . t_1 \quad (10)$$

and

$$\omega_1 = \omega_0 + \alpha . t_1 \quad (11)$$

where, according to equation (9), $\omega_1 = v_1/R$, leading to

$$t_1 = \frac{v_0 - \omega_0 R}{\alpha R - a} \quad (12)$$

Substituting the corresponding experimental values gives $t_1 \sim 0.95$ s, which is consistent with the $t_1$ value observed in Figures 3 and 4.

Finally, the shapes of the experimental plots of the linear and angular velocities are fully consistent with equations (10) and (11), where, according to equation (8), α has a positive sign.

## Final remarks

The analysis of the system's dynamics and kinematics is simple, yet conceptually rich as well as highly illustrative. The fact that the force of friction is the only acting force having a horizontal component—as can be clearly seen in the free body diagram—helps students to visualize and better understand that it must be positive in sign and therefore responsible for accelerating the wheel.

In the plots, the constant increase in the velocity of the centre of mass over the time interval where slipping occurred can be readily observed by students. The angular velocity curve plotted by *Tracker* also clearly illustrates the transition from slipping to rolling, as well as the positive sign of the angular acceleration. Comparing the experimental plots with their reference theoretical equations serves as a visual aid that helps students to reinforce the underlying physical laws.

The experiment can be carried out easily at low cost and is suitable for both indoor as well as outdoor class time. It can be conducted in small groups or used *Interactive Lecture Demonstrations* or other active learning settings. Data acquisition is hardly time consuming and the processing of experimental data is performed automatically by *Tracker*. Using the appropriate tool of *Tracker*, the concept of moving frames of reference can be introduced in an intuitive manner.

## Acknowledgements

We acknowledge financial support from grant FSED_3_2016_1_134232 (ANII-CFE). Translated from the Spanish by Eduardo Speranza.